\documentclass[twocolumn,prl,floats,showpacs]{revtex4}
\usepackage{graphicx}

\begin{document}

\title{Does the continuum theory of dynamic fracture work?}

\author{David A. Kessler}
\affiliation{Department of Physics, Bar-Ilan University, Ramat-Gan,
Israel}
\author{Herbert Levine}
\affiliation{Department of Physics, University of California, San
Diego, La Jolla, CA 92093-0319}

\date{\today}

\begin{abstract}
We investigate the validity of the Linear Elastic Fracture Mechanics
approach to dynamic fracture.  We first test the predictions in a
lattice simulation, using a formula of Eshelby for the 
time-dependent Stress Intensity Factor. Excellent agreement
with the theory is found.  We then use the same method to analyze the
experiment of Sharon and Fineberg.  The data here is not consistent
with the theoretical expectation.
\end{abstract}
\maketitle

Recently, there has been renewed interest in the physics community
concerning the problem of dynamic fracture. This interest was
kindled by experiments\cite{swinney,sharon} showing a universal transition to 
crack branching and the realization that one could not approach this
question within the confines of traditional fracture mechanics\cite{review}.
Since then, there have been other novel experimental findings\cite{balloon} as
well as further evidence of theoretical inadequacies.

In the standard approach, (see, for example, \cite{broberg})
dubbed Linear Elastic Fracture Mechanics,
one assumes that continuum elasticity is
valid everywhere outside of an microscopically sized process zone.
The instantaneous crack tip velocity is then postulated to depend
only on the singular part of the stress field obtained by solving
this macroscopic continuum problem. This singularity of the stress
field is universal in nature, up to an overall multiplicative factor,
the so-called stress-intensity factor (SIF).  Even if there is a well-defined
 relationship
between the crack velocity and the SIF, the theory cannot
predict the form of the relationship, as that depends explicitly on
physics at the scale of the process zone. As realized initially by
Slepyan\cite{slepyan}, one way to remedy this deficiency is to
model the entire system as a lattice of mass points connected by
nonlinear springs. On scales large compared to the lattice
spacing, the displacement approaches that predicted by the
continuum theory; on the scale of the crack tip, the stress field
divergence is regularized, thereby allowing for the imposition of
a physically sensible breaking criterion. This criterion is
usually in the form of a critical spring displacement, as this
allows for the possibility of analytical solutions of the
model\cite{marder+gross,kl1,k2,leo1,slepyan_new,marder_nature}.

Thus, the lattice model provides a self-consistent realization of
the basic assumption underlying engineering fracture mechanics,
the ability to separate the linear elasticity calculation from the
microscopic physics controlling the tip. 
In general, however, it is hard to test the LEFM since it is difficult 
to reliably measure the SIF in a
lattice calculation, as this requires an extremely (and impractically)
large system.
Only then is there an appreciable range of scales in which one is
sufficiently far from the crack tip that lattice effects are unimportant
and sufficiently close that the fairly weak square-root singularity 
dominates.  However, for the case of a crack accelerating from rest,
the approach of Kostrov \cite{kostrov} and Eshelby \cite{eshelby}
 provides an analytic prediction of the SIF, independent
of the details of the dynamics.  With this first-principles
determination of the SIF, it is possible to test the LEFM picture.
Specifically, we will show that in the lattice model, 
in accord with expectations, the
stress intensity factor governing the strength of the stress
singularity is the only information passed from the macroscopic
field to the process-zone dynamics of the crack tip. In fact, this
separation is quantitatively accurate even for rather small
lattices where one might have questioned the efficacy of the
continuum approach. Moreover, these results allow us to
construct a test of the theory based on the actual fracture data
presented by Sharon and Fineberg\cite{nature} for the
same case of a crack accelerating from rest. Here, however,
the data do not appear
to conform to theoretical expectations. At the end, we discuss
possible implications of this failure.

We begin with the lattice model. We work with a 2d square lattice
with unit spacing between the mass points. These masses are
coupled via both nearest neighbor and next-nearest neighbor
ideally brittle central force springs with spring constants $k_1$
and $k_2$ respectively. It is easy to show that with the choice
$k_1=2k_2=2\mu=2\lambda_{2d}$, the continuum linear elastic limit
of this model is isotropic with the aforementioned Lame'
constants; hence the Poisson ratio $\nu _{2d}=1/4$. We will assume
that the actual 3d system exhibits plane stress and so
can be approximated via a 2d system with
$$ \lambda _{2d} = \frac{2\mu \lambda}{2
\mu +\lambda} \ ;$$ the actual Poisson ratio of the material being
modeled is $\nu = \nu_{2d}/(1-\nu_{2d}) =1/3$. The dynamics
arising from this force is taken to include the possibility of
a Kelvin viscosity
term. The final equation of motion is therefore
\begin{eqnarray}
\frac {\partial^2 \vec{u} (\vec{x})}{\partial t^2} & =& \ \left( 1+
\eta \frac{\partial}{\partial t} \right) \left[  
\frac{k_1}{2} \sum_{\hat{n}\in nn}
((\vec{u}(\vec{x}+\hat{n})-\vec{u}(\vec{x}))\cdot \hat{n})\hat{n} \right.
\nonumber \\ &\ & \quad  +  \left.\frac{k_2}{2} \sum_{\hat{n}\in nnn}
((\vec{u}(\vec{x}+\hat{n})-\vec{u}(\vec{x}))\cdot \hat{n})\hat{n} 
\right]
\end{eqnarray}
As discussed in \cite{kelvin,kl1}, the damping due to nonzero
Kelvin viscosity occurs only inside the process zone if $\eta$ is
chosen $0(1)$. Finally, any bond whose length goes
above a breaking threshold, which we take to be 1, 
has its spring constant set to zero.

As discussed above, we study in detail a
finite length crack accelerating from rest. Initially, a crack is
placed along the mid-line of a sample (here between rows 0 and 1
of our lattice), extending a length $\ell_0$ from the left edge. 
The top and bottom rows
of the lattice have fixed (and opposite) displacements
$\vec u=\Delta \hat{y}$ and the lateral edges are free. 
The loading is chosen to
be just below the critical value at which the crack will start to
propagate; since this loading is a decreasing function of crack
length, a crack with one additional broken bond will in fact start
to elongate. The system is then allowed to fully relax to its
equilibrium stress state; this is accomplished via a multigrid
technique described in detail elsewhere\cite{long}. Once this is done, an
additional bond is broken by hand (i.e., its spring constant is set to zero; no
actual displacement of particles is involved) and the crack tip accelerates.
As it moves, we monitor the bonds across the crack surface $y=1/2$. When
all three bonds attached to a given point $\vec{x} = (x,1)$ break,
the crack length is deemed to have increased by one and the
velocity at that time is measured as the inverse of the time
interval since the last such event. We do not allow bonds off the
crack line to break, thereby suppressing any possible branching
instabilities. We also enforce symmetry across the crack surface, simulating
only the upper half of the lattice. Each run is characterized by the 
transverse lattice
size $W$, the initial length of the crack $\ell_0$,
damping constant $\eta$ and the driving displacement $\Delta$.
Typical data generated by this procedure for both
the undamped and highly damped cases are presented in Fig. 1.

\begin{figure}
\includegraphics[width=3.5in]{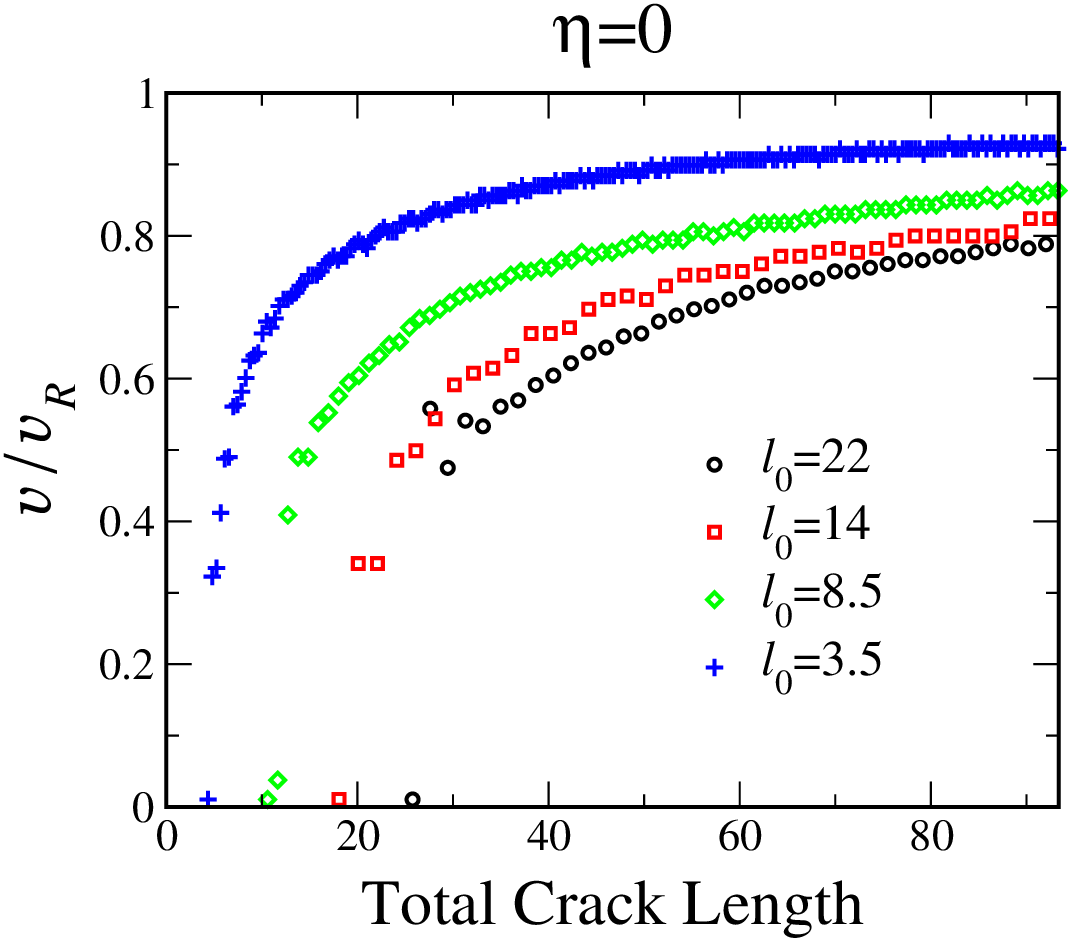}
\includegraphics[width=3.5in]{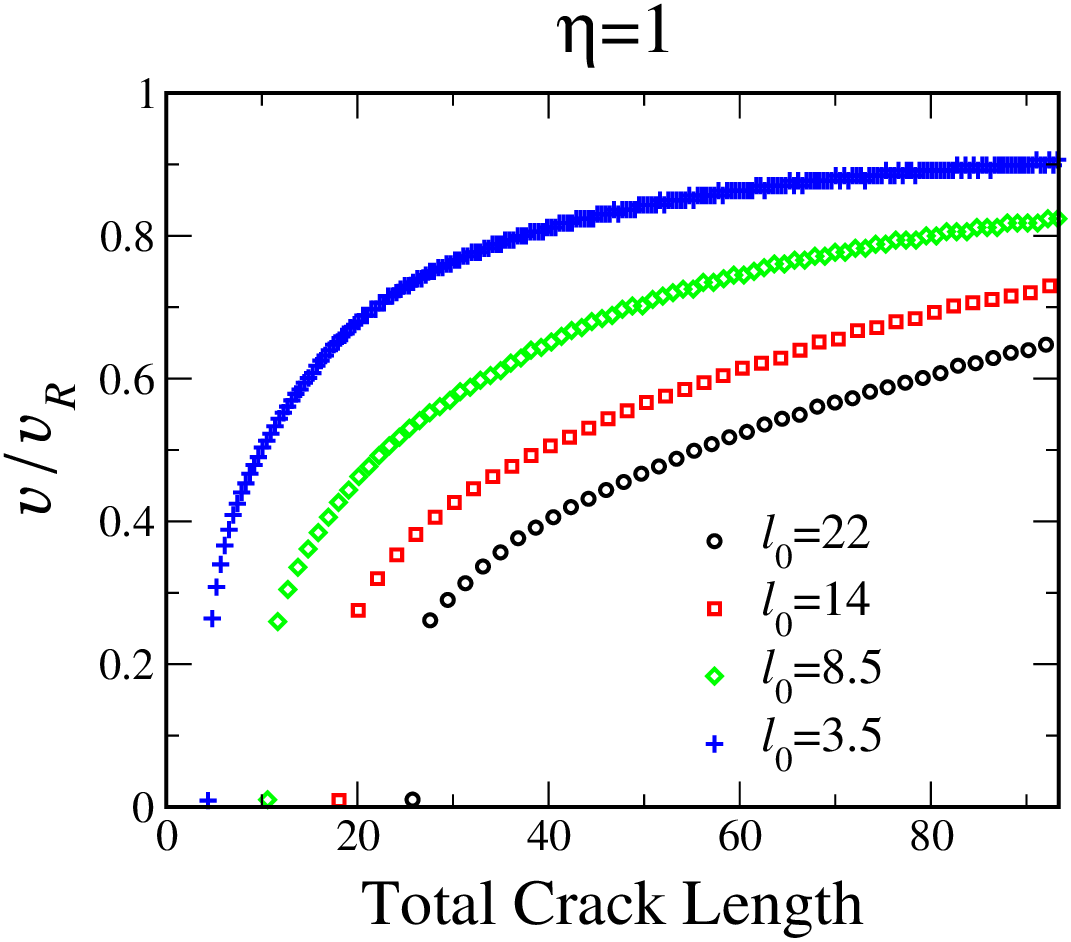}
\caption{Simulation results for velocity vs. crack length for various
initial seed cracks.  The system widths, $W$, in lattice units
are 239, 219, 415, and 1004
for the four runs for decreasing $\ell_0$. 
For later convenience, all lengths in the graph are scaled by a factor 440/$W$.
Velocities are normalized to the Rayleigh velocity. }
\label{fig1}
\end{figure}

According to the classic calculation by Eshelby \cite{eshelby},
the stress
intensity factor, $K$, or equivalently the Eshelby $B$ factor (which is a
constant multiple of $K$), at the progressing crack tip for a system that
starts in equilibrium can, up to a certain time, be written as a product of 
two factors:
\begin{equation}
B_I (t) \ = \ A(v(t))) \int_{\ell_0} ^{\ell (t)} dx \frac{
\sigma _{eq}^{yy} (x)} {\sqrt{x- \ell_0}}
\end{equation}
Here $\sigma_{eq}^{yy}$ is the normal stress on the mid-line $y=1/2$ as
found from the equilibrium stress field; it diverges of course
near the edge of the equilibrium crack $x = \ell_0$ with a static
stress intensity factor. The first factor $A(v)$ depends only on
the {\em instantaneous} velocity at time $t$ and the second factor
(which we will refer to as $B_0$) depends on time only through the {\em
instantaneous} crack length $\ell (t)$. This equation holds true as
long as sound waves reflected from the boundaries have not interacted with the
crack tip.  There are two boundaries one must be concerned with,
the lateral boundaries at $y=\pm W/2$ and the edge at $x=0$.  We take
care that all our
data come from times before these interactions occur.

Now, if the true
microscopic breaking events only couple to the macroscopic field
via $B_I$, we expect that there will be some fixed relationship
between the tip velocity and this number. Usually, this
relationship is considered to arise due to energy conservation and
is therefore written as an equality of the energy flux into the
tip and the energy necessary to break bonds along a unit length of
crack
\begin{equation}
\Gamma (v) \ = \ f(v) B_I ^2
\end{equation}
Here $f$ is a complicated but explicit function and $\Gamma$ is
the breaking energy. In fact, $\Gamma$ can be determined via the
dependence of steady-state cracks on the driving load and
thereafter used for the accelerating crack case. For our purposes,
this hypothesis leads to the existence of a universal
relationship, independent of $W$, $\Delta$ and $\ell_0$ (but dependent on
$\eta$) , between the measured velocity $v(t)$ and the Eshelby
function $B_0(\ell (t))$.

To test this strong prediction of LEFM, we use our
lattice model simulations discussed above. We calculate the
integral in Eq. (2) by replacing the continuum stress with its
lattice analog (defining $\vec u \equiv (u,v)$)
\begin{eqnarray}
\sigma _{eq}^{yy} (x) & \  =  \ - k_1 v_{eq}(x,0) \ - \ \frac{1}{2} k_2
\left[ u_{eq}(x-1,0) - \right. \nonumber \\ & - u_{eq}(x+1,0) +
v_{eq}(x-1,0) +v_{eq} (x+1,0) \left. \right]
\end{eqnarray}
The fields that enter are the equilibrium fields present before
the crack tip begins to move. The integral is then replaced by a
sum, taking care to resolve the singularities in the integrand.
This yields a function $B_0 (\ell ; W, \ell_0 )$ which can be
plotted versus the velocity. The results of this exercise are
presented in Fig. 2, demonstrating almost perfect data collapse
even for fairly small transverse sizes $W$.  

\begin{figure}
\includegraphics[width=3.5in]{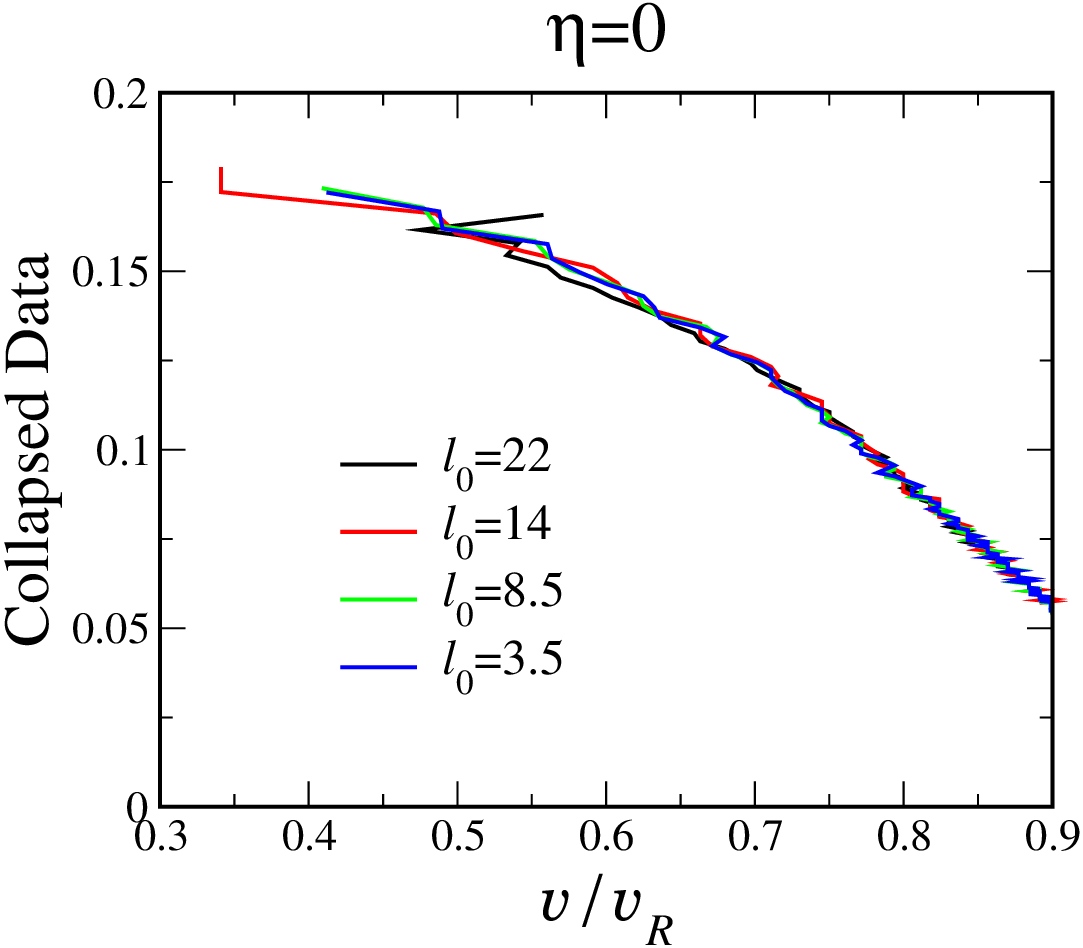}
\includegraphics[width=3.5in]{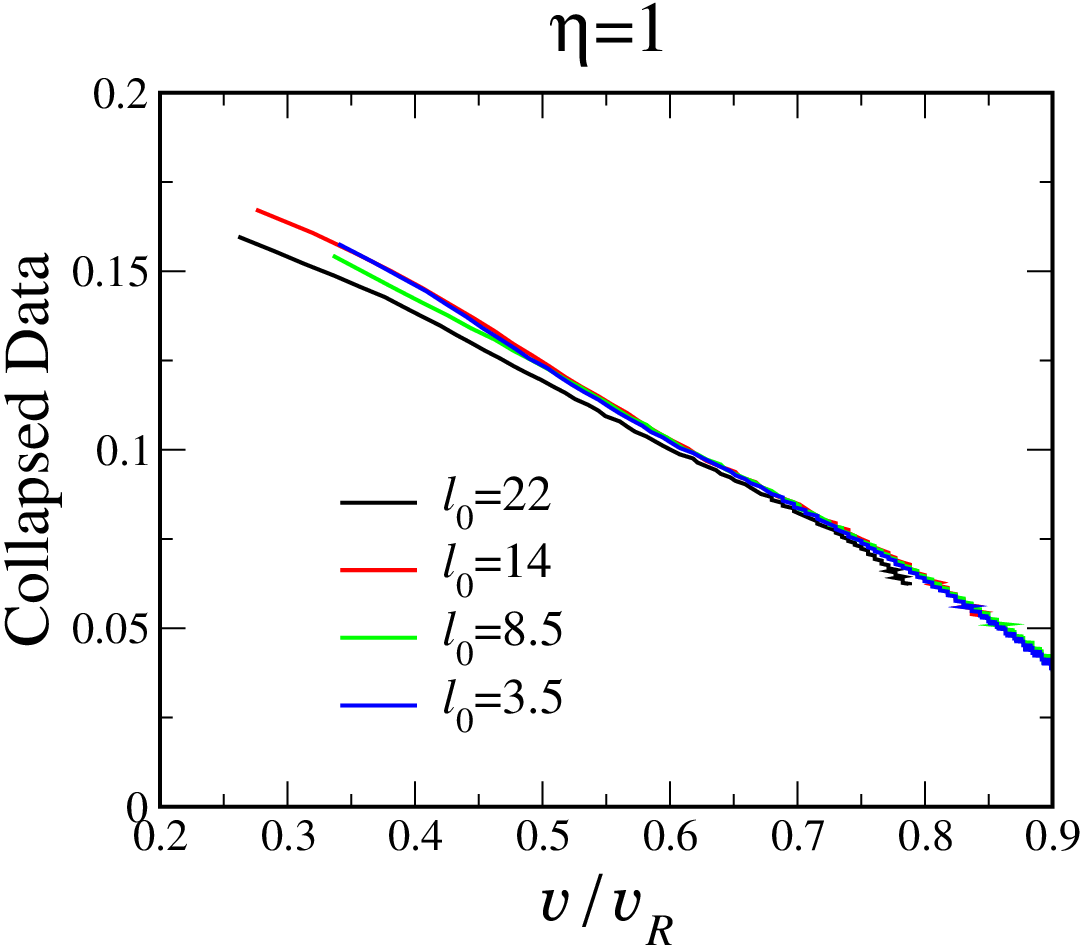}
\caption{Eshelby data collapse for $\eta=0, 1$ for the runs shown in Fig. 1.
Notice the slight systematic dependence on $\ell_0$ for the $\eta=1$ data, 
presumably arising
from the larger process zone in this case \protect{\cite{k2}}.}
\end{figure}

So, the lattice model does indeed serve as an instantiation of the
LEFM picture developed by the fracture community for straight
accelerating cracks. But now, we can use the Eshelby $B_0$
function calculated as just described to process actual crack
velocity/length data from the experiments on PMMA carried out by
Sharon and Fineberg. This data comes from a protocol very similar
to what we used above in which the loading is set to a
point where a touch of razor blade will cause the rapid elongation
of a previously created notch in the material. For PMMA, the
elastic constants are $\lambda = 2680 MPa$, $\mu =  MPa$, giving
$\nu = .35$. Going to the plane stress case leads to an effective
Poisson ratio $\nu _{2d} \simeq .259$ which is very close to the
value forced upon us by the restriction to n.n. and n.n.n. central
forces. We did check that generalizing our model to include
bond-bending springs and thereby obtaining the precise value of
the Poisson ratio did not alter in any way the results which we
will now present. 

In fact, we designed our simulation to mimic as closely as possible
the experimental situation.  Thus, not only were the elastic constants
chosen appropriately, the aspect ratio was also fixed to that of the
experiment ( a width of 440 mm. and a length of 380 mm.).  The lengths
of the initial seed cracks were also chosen to be in the same ratio
to the width as in the experiment.  This constraint fixed our choice
of the actual width in lattice units, as we wanted the initial crack
to be large (of order 10) on the lattice scale.

Fig. 3 shows our attempt to verify the continuum hypothesis for
the PMMA data. 
In our opinion, the data are quite convincingly
inconsistent with the universality of the $v-B_0$ relationship.
Note that this is the opposite conclusion from that reached by the
experimenters themselves, as discussed in Ref. \cite{nature}.
There, the exact Eshelby function was replaced by an approximate
form based on the {\em static} stress intensity factor for a crack
of length $\ell (t)$. Note that these functions while close for
$\ell \approx \ell_0$, become quite different for larger length cracks.
The static stress intensity factor saturates for $\ell \gg W/2$, whereas
the exact Eshelby function increases linearly in this regime.
We have checked that the use of this
approximation happens to push the data into closer agreement with
the theory and could lead to a mistaken impression of data
collapse. This does not happen with the exact expression and
conversely the approximate form does not lead to a very good data
collapse for the simulation study discussed above.  The same situation
obtains for the Sharon-Fineberg measurements on fracture in glass (results
not shown).
There the data collapse is, if anything, even worse.

\begin{figure}
\includegraphics[width=3.35in]{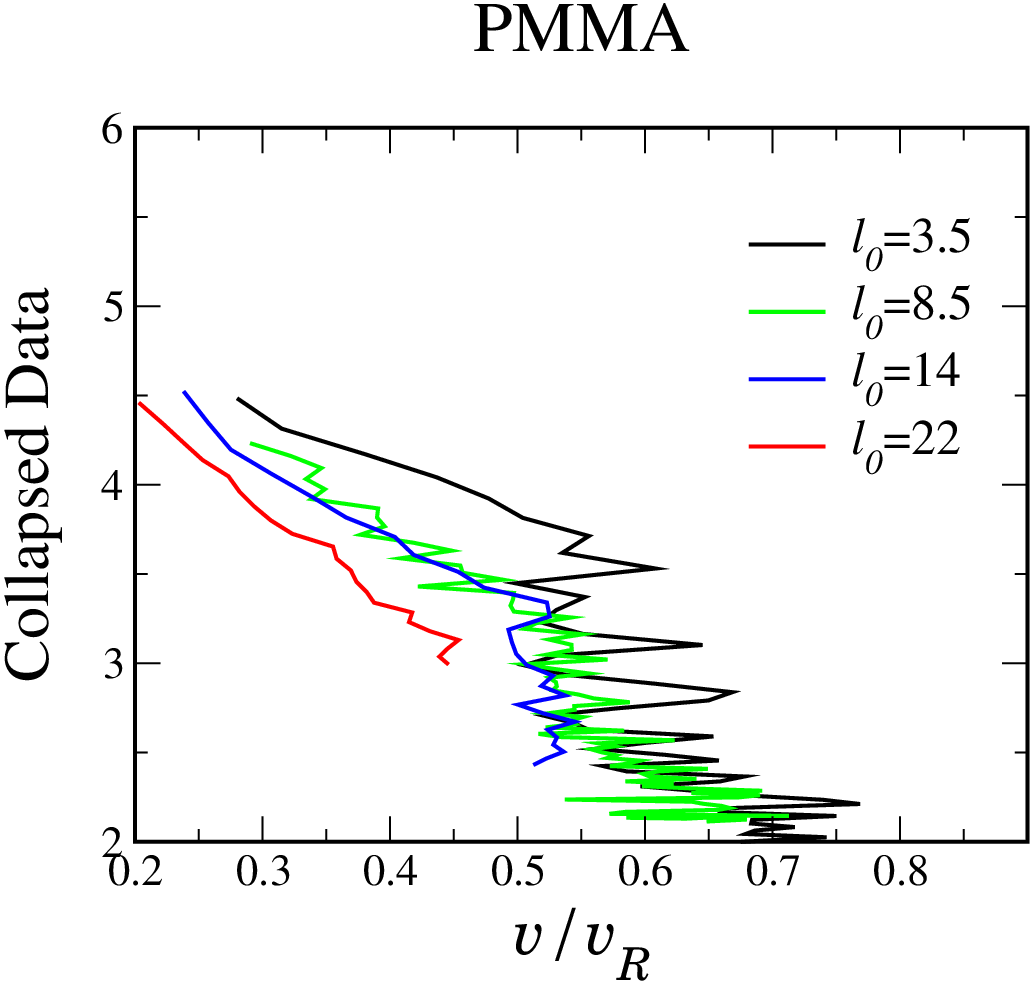}
\caption{Attempted data collapse for the Sharon-Fineberg PMMA experiment. 
Initial crack lengths are given in mm.}
\end{figure}

It should be noted that the agreement between the lattice theory
and LEFM is very dependent on our absolute suppression of off-axis
cracking.  If all bonds are allowed to break, above a critical
velocity the crack no longer
propagates in a straight line, and the assumptions implicit in the
Eshelby calculation of the instantaneous SIF break down completely.  At
that point, there is no longer any way to reliably measure the SIF
and thus test LEFM.  Needless to say, LEFM in its simplest formulation
does not predict the direction of branching, and thus cannot address
to the post-instability dynamics of the crack.  The effects of the
instability in the experiment (which appear to exhibit different dynamics
than in the lattice model in this regime) are clearly visible at later
times (larger velocities), but cannot explain the apparent failure of the
data collapse at smaller velocities.

The question remains how to explain this failure of LEFM in the
Sharon-Fineberg experiment.  Putting aside for the moment the
obvious possibility of systematic errors in the determination of the
velocity or position of the crack, the only explanation that suggests
itself is that the process zone is for some reason very large.
The monotonic dependence on $\ell_0$, similar to,though much larger than 
that seen in the
$\eta=1$ simulation, is consistent with this hypothesis.
This might be due either to the polymeric nature of the microstructure
of the polymer system, the heterogeneity of the small-scale structure,
or to the large dissipation in the system.
However, given the success of LEFM on our relatively small systems,
it is not easy to accept this as a sufficient explanation.  Further
experiments on other systems are clearly necessary for an unraveling
of this conundrum.

\begin{acknowledgments}
The authors wish to thank J. Fineberg and E. Sharon for providing the
raw data from their experiment, and for extensive
discussions. The work of DAK is supported in part by the Israel
Science Foundation.  The work of HL is supported in part by the
NSF, grant no. DMR-0101793.
\end{acknowledgments}

\end{document}